\documentclass[twocolumn,prb,showpacs,superscriptaddress]{revtex4}

\usepackage{psfrag}
\usepackage{graphicx}
\usepackage{amsmath}
\usepackage{amssymb}

\newcommand{\bra}[1]{\langle #1 |}
\newcommand{\ket}[1]{| #1 \rangle}
\newcommand{\braket}[2]{\langle #1 | #2 \rangle}

\begin{document}
\title{First-principles many-body calculations of electronic conduction in thiol- and amine-linked molecules}

\author{M. Strange}
\affiliation{Center for Atomic-scale Materials Design, Department of
Physics, Technical University of Denmark, DK - 2800 Kgs. Lyngby, Denmark}
\affiliation{Departments of
Physics and Chemistry, University of Jyvaskyla, Jyvaskyla, Finland}
\author{C. Rostgaard}
\affiliation{Center for Atomic-scale Materials Design, Department of
Physics, Technical University of Denmark, DK - 2800 Kgs. Lyngby, Denmark}
\author{H. H\" akkinen}
\affiliation{Departments of
Physics and Chemistry, University of Jyvaskyla, Jyvaskyla, Finland}
\date{\today}
\author{K. S. Thygesen}
\email[]{thygesen@fysik.dtu.dk}
\affiliation{Center for Atomic-scale Materials Design, Department of
Physics, Technical University of Denmark, DK - 2800 Kgs. Lyngby, Denmark}

\date{\today}

\begin{abstract}
  The electronic conductance of a benzene molecule connected to gold
  electrodes via thiol, thiolate, and amino anchoring groups is
  calculated using nonequilibrium Green functions in combination with
  the fully selfconsistent GW approximation. The calculated conductance of
  benzenedithiol and benzenediamine is five times lower than predicted by standard density functional theory (DFT) in very good agreement
  with experiments.  In contrast, the widely studied benzenedithiolate
  structure is found to have a significantly higher conductance due to the unsaturated sulfur bonds.
These findings suggest that more complex gold/thiolate structures where the thiolate anchors are chemically
  passivated by Au adatoms are responsible for the measured conductance. Analysis of the
  energy level alignment obtained with DFT, Hartree-Fock and GW
  reveals the importance of self-interaction corrections (exchange) on
  the molecule and dynamical screening at the metal-molecule
  interface. The main effect of the GW self-energy is to renormalize
  the level positions, however, its influence on the shape of
  molecular resonances also affects the conductance.
Non-selfconsistent G$_0$W$_0$ calculations,
  starting from either DFT or Hartree-Fock, yield conductance values
  within 50\% of the selfconsistent GW results.
\end{abstract}
\maketitle

\section{Introduction}
The problem of first-principles calculation of electronic
conduction in molecular systems is of longstanding interest. Over
the last decade, advances in experimental techniques have allowed for
fundamental studies of electron transport through few or even
individually contacted molecules. The transport mechanisms observed in
molecular junctions range from ballistic\cite{smit02} over
off-resonant tunneling\cite{venkataraman,song,xiao,tsutsui,kiguchi} to the
strong correlation Kondo and Coulomb blockade
regime\cite{kubatkin,park} to vibration assisted hopping\cite{choi}.
The former two belong to the phase-coherent transport regime
characteristic of relatively short molecules (as opposed to molecular
wires\cite{choi}) with good "chemical" contact to the electrodes and is the main focus of the present work.

The lack of control over the atomistic details of the metal-molecule
interface introduces a strong statistical element in measurements on
single-molecule junctions which masks the relation between atomic
structure and measured conductance. For example, published conductance
values for the gold/benzenedithiol model system vary by several orders
of magnitude, although recent independent studies of this system seem
agree on a value around $0.01G_0$
($G_0=2e^2/h$)\cite{song,xiao,tsutsui,kiguchi}. It is, however, not
clear which structure is responsible for this "typical" conductance
and recent experimental and theoretical work point to complex
gold/thiolate structures involving two molecules bonding the same Au
adatom\cite{ref18,ulstrup,voznyy,ref19,ref20,ref21}.  Scanning
tunneling microscope experiments in solution have shown that the use
of amino rather than thiol anchoring groups leads to more well defined
junction properties\cite{venkataraman} and a conductance around (or just below)
$0.01G_0$ has also been reported for the gold/benzenediamine junction\cite{venkataraman,kiguchi}. This was,
however, not confirmed by independent break junction experiments in
vacuum\cite{ruitenbeek}.

The uncertainties related to the junction atomic structure renders
theoretical benchmark calculations more important and more challenging
at the same time. Here progress has been hampered by the inability of
conventional density functional theory (DFT) methods, which for long have
been the workhorse and the state of the art for quantum transport
calculations\cite{brandbyge,xue}, to reproduce the conductance
measured for even the simplest molecular tunnel junctions\cite{magnus,evers}.  As a
consequence, the most succesful studies have focused on qualitative
trends in the dependence of conductance on e.g. molecular
conformation\cite{mischenko}, molecular length\cite{hybertsen}, side
group functionalizations\cite{mowbray}, or have focused on properties
independent of the numerical value of the conductance like molecular
vibrations\cite{okabayashi,raman}. The main shortcoming of the DFT
approach has been attributed to its band gap problem, i.e. the fact
that DFT tends to underestimate energy gaps\cite{godby1}, and therefore
overestimates the conductance. Attempts to overcome this problem 
within a single-particle framework have mainly focused on self-interaction correction schemes\cite{burke,sanvito,baranger}.

The well known success of the many-body GW method\cite{hedin}
for quasiparticle (QP) bandstructure calculations has recently inspired its application
to (simplified) transport problems\cite{thygesen_jcp,gw_prb,darancet,thygesen_prl,leeuwen_trans,leeuwen_trans2,spataru1,spataru2}.
The fact that the GW approximation succesfully describes systems with
highly diverse screening properties ranging from metals\cite{barth} over
semiconductors\cite{usuda_hamada,ku,schilfgaarde_prb,Shishkin2006,rinke,orr,kresse} to molecules\cite{rostgaard,stan_eulett}
is essential for a correct description of metal-molecule interfaces where the electronic character changes from metallic to insulating over a few angstrom. In particular, screening by the metal electrons can have large influence on the QP energies of the adsorbed
molecule\cite{neaton,juanma,thygesen_image,kaasbjerg} -- an effect completely missed by both local and hybrid density functionals\cite{juanma}. This has recently
motivated the use of semi-empirical schemes for correcting the DFT
eigenvalues by a scissors operator prior to transport
calculations\cite{quek,mowbray,quek2}. While such schemes can be justified
for weakly coupled molecules, they become uncontrolled  in the relevant
regime of covalently bonded molecules where the screening effects mix with
charge transfer and hybridization\cite{thygesen_image}.

In this work we combine nonequilibrium Green function methods for electron
transport with the fully selfconsistent GW approximation for exchange
and correlation to establish a theoretical benchmark for the
electronic structure and conductance of gold/1,4-benzenedithiolate
(BDT), -benzenedithiol (BDT+H) and -benzenediamine (BDA) molecular
junctions.  We find a conductance of $0.0042G_0$ for BDA and
$0.010G_0$ for BDT+H in very good agreement with experimental data. In
comparison, the conductance obtained from DFT is about five times
higher while non-selfconsistent G$_0$W$_0$ calculations produce
conductances within 50\% of the selfconsistent GW result. We argue
that the BDT+H structure can be viewed as a simple model of recently
proposed RS-Au(I)-SR gold/thiolate structures involing two molecules attached to
the same Au adatom. The conductance of a simple BDT molecule between Au(111) surfaces is
predicted to be on the order of $1G_0$ by both DFT, Hartree-Fock and
GW. The origin of the high conductance is due to an unsaturated sulfur $p$
orbital with energy just below the Fermi energy. In the BDT+H and  
RS-Au(I)-SR structures, the sulfurs are fully passivated and the $p$-orbital moves away from $E_F$ leading to an effective decoupling of the C$_6$H$_4$ moiety from the gold electrodes. 
We find that the main effect of the GW self-energy is to shift the molecular levels and can be modelled by a simple scissors
operator. However, the energy dependence of
the GW self-energy may also affect the shape of the transmission resonances
and this can change the conductance by almost a factor of two.

Most implementations of the GW method invoke one or several technical
approximations like the plasmon pole approximation, neglect of
off-diagonal matrix elements in the GW self-energy, analytic
continuations from the imaginary to the real frequency axis, neglect
of core states contributions to the self-energy, neglect of
self-consistency. The range of validity of these approximations has
been explored for solid state systems by a number of
authors\cite{barth,usuda_hamada,ku,schilfgaarde_prb,Shishkin2006,rinke},
however, much less is known about their applicability to molecular and
metal-molecule systems. For this reason our implementation of the GW
method avoids all of these approximations and as such represents an exact
treatment of the GW self-energy within the space of the employed
atomic orbital basis set.

Although we compare our results to experimental data and discuss them
in relation to the possible atomic structure of the junctions, we stress that the main purpose of this study is the
benchmarking of quantum transport calculations for specific, idealized junction with particular focus on the role 
of electronic correlation effects.

\section{Method}\label{sec.method}
We consider a quantum conductor consisting of a molecule connected to
left (L) and right (R) electrodes. We shall assume that outside a certain
region containing the molecule and part of the electrodes (the "extended molecule"), the
electron-electron interactions can be described by a mean field potential.  The current
through the molecule is then given by\cite{ref30,nonorthogonal}
\begin{equation}\label{eq.eq0}
I=\frac{i}{4\pi}\int \text{Tr}[(\Gamma_L-\Gamma_R)G^{<}+(f_L\Gamma_L-f_R\Gamma_R)(G^r-G^a)]\text{d}E
\end{equation}
where the energy dependence of all quantities has been suppressed for simplicity. In this equation $G$ is the Green function 
matrix of the extended molecule evaluated in a localized basis, $\Gamma_{L/R}=
i[\Sigma^{r}_{L/R}-\Sigma^a_{L/R}]$ is the coupling strength between
the extended molecule and the left/right electrode, and $f_{L/R}$ are
the Fermi Dirac distribution functions of the two leads. Our
implementation applies to the general case of a finite bias voltage, but
in this work we focus on the zero bias conductance which can be
expressed in terms of the transmission function\cite{ref30,nonorthogonal}
\begin{equation}\label{eq.eq1}
T(E) =\text{Tr}[G^r(E)\Gamma_L(E)G^a(E)\Gamma_R(E)]
\end{equation}
as $G=G_0T(E_F)$.  This formula was originally derived for
non-interacting electrons, but is in fact valid for interacting
electrons in the low-bias limit\cite{godby}. We have verified that this is indeed
fulfilled to high numerical accuracy in our GW calculations by comparing the conductance
obtained from Eq. (\ref{eq.eq0}) for small finite voltages with
$T(E_F)$ evaluated in equilibrium.

The
retarded Green function of the extended molecule is calculated from
\begin{eqnarray}\nonumber
G^r(E)=[(E+i\eta) S - H_0 + V_{xc} - \Delta
V_{\text{H}}[G]\\\label{eq.gr}
-\Sigma_{L}(E)-\Sigma_{R}(E)-\Sigma_{xc}[G](E)]^{-1}
\end{eqnarray}
where $\eta$ is a positive infinitesimal and
\begin{eqnarray}
S_{ij}&=&\langle \phi_i|\phi_j\rangle\\
H_{0,ij}&=&\langle \phi_i|-\frac{1}{2}\nabla^2+v_{\text{ion}}(\bold r)+v_{\text{H}}(\bold r)+v_{xc}(\bold r)|\phi_j\rangle\\
V_{xc,ij}&=&\langle \phi_i|v_{xc}(\bold r)|\phi_j\rangle
\end{eqnarray}
denote the overlap matrix, Kohn-Sham Hamiltonian and
exchange-correlation potential, respectively. The matrices are
evaluated in terms of a basis consisting of numerical atomic
orbitals\cite{GPAW-LCAO}, and are obtained from a DFT supercell
calculation performed with the real-space projector augmented wave
method GPAW\cite{ref29}. The electrode self-energies $\Sigma_{L/R}$
are obtained from the Kohn-Sham Hamiltonian of a bulk DFT calculation
using standard techniques\cite{ref10}. 

The boundary conditions in the
plane normal to the transport direction enter only via the electrode
self-energies which are constructed from the electrode surface Green
function\cite{ref10}. The latter should represent an
infinite surface but is here approximated by that of a periodic supercell with
$4\times 4$ Au atoms in the surface plane. We have found that this is
a very good approximation when the surface Green function of the
$4\times 4$ cell is evaluated at a general $k_{\perp}$-point (we use $k_{\perp}=(0.1,0.2)$ in
coordinates of the surface Brillouin zone basis vectors). Using high symmetry points, in particular the Gamma point, can be problematic\cite{k-point}.

The term $\Delta
V_{\text{H}}$ is the deviation of the Hartree potential from the groundstate DFT Hartree potential contained in $H_0$, 
\begin{equation}\label{eq.hartree}
\Delta V_{\text{H},ij}= 2\sum_{kl}v_{ij,kl}(\varrho_{kl}[G]-\varrho^0_{kl}[G_0])
\end{equation}
In this expression $\varrho=-i\int G^<(E)dE$ and $\varrho^0=-i\int G^<_0(E)dE$ are the interacting and Kohn-Sham density matrices, respectively, and   
\begin{equation}\label{eq.cou}
v_{ij, kl} = \int \int \frac{\phi_i^*(\bold r) \phi_j(\bold r) \phi_k(\bold r') \phi_l^*(\bold r')}{
|\bold r-\bold r'|} \mathrm{d}\bold r \mathrm{d}\bold r'
\end{equation}
is the bare Coulomb interaction in the atomic orbital basis. The
factor 2 is due to spin. Ref. \onlinecite{Walter2008} describes how
the all electron Coulomb elements are obtained within the PAW
formalism. The last term in Eq. (\ref{eq.gr}) is the many-body
exchange-correlation self-energy which in this work can be either the
bare exchange potential or the GW self-energy. We note that setting
$\Delta V_{\text{H}}=0$ and $\Sigma_{xc}=V_{xc}$ in Eq. (\ref{eq.gr})
we obtain the Kohn-Sham Green function, $G_0$.

\subsection{GW self-energy}\label{sec.gw}
The (time-ordered) GW self-energy is given by
\begin{equation}\label{eq.gw}
\Sigma_{\text{GW}, ij}(\tau) = i\sum_{kl} G_{kl}(\tau^+)W_{ki,lj}(\tau)
\end{equation}
where $W$ is the screened interaction and the indices $i,j,k,l$ run
over the atomic basis functions of the extended molecule. The screened interaction is
calculated in the frequency domain as the matrix product
$W(\omega)=\epsilon^{-1}(\omega)v$ with the dielectric function,
$\epsilon(\omega)=1-vP(\omega)$, evaluated in the random phase
approximation. The irreducible density reponse function is computed in
the time domain,
\begin{equation}\label{eq.p}
P_{ij, kl}(\tau) =-2 i G_{ik}(\tau)G_{lj}(-\tau).
\end{equation}
where the factor 2 accounts for spin. Setting $P=0$ yields the Hartree-Fock
approximation which thus corresponds to complete neglect of screening or equivalently complete neglect of correlations.
Note that Eqs. (\ref{eq.gw}-\ref{eq.p}) involve time-ordered quantities defined on the Keldysh time contour. For completeness we provide the expressions for the real time components in Appendix \ref{sec:GWselfenergy} and refer the reader to Ref. \onlinecite{gw_prb} for more details.

\begin{figure}[!h]
    \includegraphics[width=0.8\linewidth,angle=0]{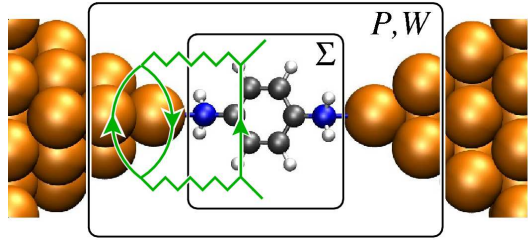}
    \caption[cap.Pt6]{\label{fig0} (Color online) Schematic of a molecular junction with the "molecule" and "extended molecule" regions indicated (small and large box, respectively). The GW self-energy ($\Sigma$) is calculated on the molecule while the polarization ($P$) and screened interaction ($W$) are evaluated for the extended region to ensure proper treatment of non-local screening effects. The electrodes and electrode-molecule coupling is described at the DFT level. A second order Feynman diagram is shown.}
\end{figure}

As explained above, the matrices $P$, $W$, and $\Sigma_{\text{GW}}$ are
calculated for the extended molecule. On the other hand it is clear
that $W$, and thus $\Sigma_{\text{GW}}$, should have contributions from
polarization diagrams outside this region, see Fig. \ref{fig0}.
Physically these diagrams describe the potential acting on an electron propagating on the molecule due to the polarization that it induces in the lead. For this reason the self-energy will not be fully
converged at the ends of the extended molecule region. To overcome this
problem we only use the part of $\Sigma_{\text{GW}}$ corresponding to the
molecule and replace the remaining parts of the xc self-energy of the
extended molecule by the DFT xc-potential. Symbolically,
\begin{equation}\label{eq.sigma2}
\Sigma_{xc}(E)=\left(
                \begin{array}{ccc}
                v_{xc} & v_{xc} & v_{xc} \\
                v_{xc} & \Sigma_{\text{GW}}(E) & v_{xc} \\
                v_{xc} & v_{xc} & v_{xc} \\
                \end{array}
               \right),
\end{equation}
We stress that although we only include $\Sigma_{\text{GW}}$ on the
molecule, the interactions between electrons on the molecule and
electrons in the electrodes (leading e.g. to image charge renormalization of
the molecular levels) are included via diagrams of the form shown in
Fig. \ref{fig0}. We also note that the form (\ref{eq.sigma2}) implies that all
metal atoms, both those inside and outside the extended molecule, are
consistently described at the same (DFT) level. For the size of the extended
molecule we have found it sufficient to include the gold atoms which
are nearest neighbors to the sulfur or nitrogen atoms, i.e two gold
atoms for the tip structures and six for the flat structure depicted
in Fig. \ref{fig1}, see Appendix \ref{app3}.  We expect that this rather local screening
response is special for covalent metal-molecule bonds.

\subsection{Time/frequency dependence}
The time/frequency dependence of $G$, $P$, $W$, and
$\Sigma_{\text{GW}}$ is represented on a uniform grid ranging from
$-200$ to $200$ eV with a grid spacing of $\Delta \omega=0.01$eV. We
have verified that the results are coverged with respect to both the
size and spacing of this grid. Fast Fourier transform is used to
switch between energy and time domains to avoid convolutions. The
calculations are parallelized both over basis functions and over the
time/frequency grid points. One should always have $\eta\geq \Delta
\omega$ to ensure proper representation of possible bound states.
However, we have found that the conductance, and more generally the
transmission function at any given energy, can be linearly
extrapolated to the $\eta=0^+$ limit. This extrapolation has been
performed for all the results presented in this work.  

The memory requirements for the GW calculations (defined mainly by the size of the $P$ and $W$ matrices) are
approximately a factor 50 larger than for a corresponding DFT calculation. The GW calculations for the benzene
junctions considered in the present work were performed in parallel on 100-400
cores and took about 2 hours per selfconsisistency iteration. In
comparison a DFT calculation for the same system took around 5 hours
on 8 cores.

\subsection{Product basis}
The calculation of all of the Coulomb matrix elements, $v_{ij,kl}$, is
prohibitively costly for larger basis sets.  Fortunately the matrix is
to a large degree dominated by negligible elements. To systematically
define the most significant Coulomb elements, we use the product basis
technique of Aryasetiawan and Gunnarsson
\cite{aryasetiawan}. In this approach, the pair orbital
overlap matrix
\begin{equation}
  \label{eq:pairorb-overlap}
  S_{ij,kl} = \langle n_{ij}|n_{kl}\rangle,
\end{equation}
where $n_{ij}(\bold r)=\phi_i^*(\bold r)\phi_j(\bold r)$
is used to screen for the significant elements of $v$.

The eigenvectors of the overlap matrix Eq. \eqref{eq:pairorb-overlap}
represents a set of ``optimized pair orbitals'' and the eigenvalues
their norm. Optimized pair orbitals with insignificant norm must also
yield a reduced contribution to the Coulomb matrix, and are omitted in
the calculation of $v$. We have found that the basis for $v$ can be limited to optimized pair
orbitals with a norm larger than $10^{-5} a_0^{-3}$ without sacrificing accuracy. This gives a
significant reduction in the number of Coulomb elements that needs to
be evaluated, and it reduces the matrix size of $P(\omega)$ and
$W(\omega)$ correspondingly, see Appendix \ref{sec:GWselfenergy}.

\subsection{Valence-core exchange}
Since both core and all-electron valence states are available in the
PAW method, we can evaluate the contribution to the valence exchange
self-energy coming from the core electrons. As the density matrix is
simply the identity matrix in the subspace of atomic core states, this
valence-core exchange reads
\begin{equation}
  \label{eq:valence-core}
\Sigma_{x,ij}^{\text{core}}=  -\sum_n^\text{core} v_{in,jn},
\end{equation}
where $i, j$ represent valence basis functions and $n$ represent
atomic core states. This contribution is added to $\Sigma_{\text{GW}}$
describing the valence-valence interactions. We limit the inclusion of
valence-core interactions to the exchange potential, neglecting it in
the correlation. This is reasonable, because the polarization bubble,
$P$, involving core and valence states will be small due to the large
energy difference and small spatial overlap of the valence and core
states. In general we have found that the effect of
$\Sigma_{x,ij}^{\text{core}}$ on molecular energy levels can be up to
1 eV\cite{rostgaard}. For the benzene-like molecules considered in
this work the effect is generally less than 0.4 eV for the frontier
orbitals.

\subsection{Self-consistency}
Since $\Sigma_{\text{GW}}$ and $\Delta V_\text{H}$ depend on $G$, the Dyson
equation (\ref{eq.gr}) must be solved self-consistently in
conjunction with the self-energies. In practice, this self-consistency problem
is solved by iteration. We have found that a linear mixing of the Green functions,
\begin{equation}
G^{n}_{\text{in}}(E)=(1-\alpha)G^{n-1}_{\text{in}}(E)+\alpha G^{n}_{\text{out}}(E)
\end{equation}
with $\alpha=0.15$ generally leads to selfconsistency within 20-30 iterations.

Fully selfconsistent GW calculations are not standard, and in fact
only few previous calculations of this type have been
reported\cite{rostgaard,barth,stan_eulett}. Conventional GW
bandstructure calculations typically apply a one-shot technique where
the self-energy is evaluated with a non-interacting Green function,
$G_0$, usually obtained from DFT\cite{hybertsen2}. In comparison, the
selfconsistent approach has the immediate advantage of removing the
$G_0$-dependence, i.e. it leads to a unique solution. 

For an approach, like the present, where the chemical potential is
fixed by the external boundary conditions, some kind of
selfconsistency (though not necessarily the full GW selfconsistency
employed here) is essential to ensure charge neutrality. This is
particularly important for cases where a molecular resonance lies
close to the chemical potential. A shift in the energy of such a resonance could lead
to a large change in its occupation. In a selfconsistent calculation
this shift would be counterbalanced, mainly by a change in the Hartree
potential. On the other hand, the one-shot approach does not account
for this effect and unphysical level alignments could occur as a
result.

Finally, the fully selfconsistent GW approximation is a conserving
approximation in the sense of Baym\cite{baym62}. This becomes
particularly important in the context of transport where it ensures
that the continuity equation is satisfied\cite{baym62,gw_prb}. We
mention that the recently introduced quasi-selfconsistent GW method
(not to be confused with the fully selfconsistent GW approximation
used here), in which $G_0$ is chosen such as to mimick the interacting
Green function as closely as possible, have shown that selfconsistency
in general improves the band gaps of semiconductors as compared to
standard one-shot calculations.\cite{schilfgaarde}

\section{Results}
In this section we discuss the results of our selfconsistent GW
calculations for the electronic structure and conductance of the
prototypical gold/1,4-benzenedithiolate (BDT), -benzenedithiol (BDT+H),
and -benzenediamine (BDA) molecular junctions. We argue that the thiol
structure can be considered as a simple model for more complex
gold/thiolate structures which have been proposed recently\cite{ref21} but which are currently too large to be treated satisfactorily at the GW level. The
transport results are rationalized by considering the alignment of
molecular energy levels in the junction. Here we show that both DFT
and Hartree-Fock provide quantitatively and qualitatively wrong results by predicting a gap
opening rather than reduction when the molecule is attached to electrodes. 
Finally, we investigate to
what extent the GW results can be reproduced by a simple scissors
operator applied to the DFT Hamiltonian.

\subsection{Junction geometries}
\begin{figure*}[!t]
    \includegraphics[width=1.\linewidth,angle=0]{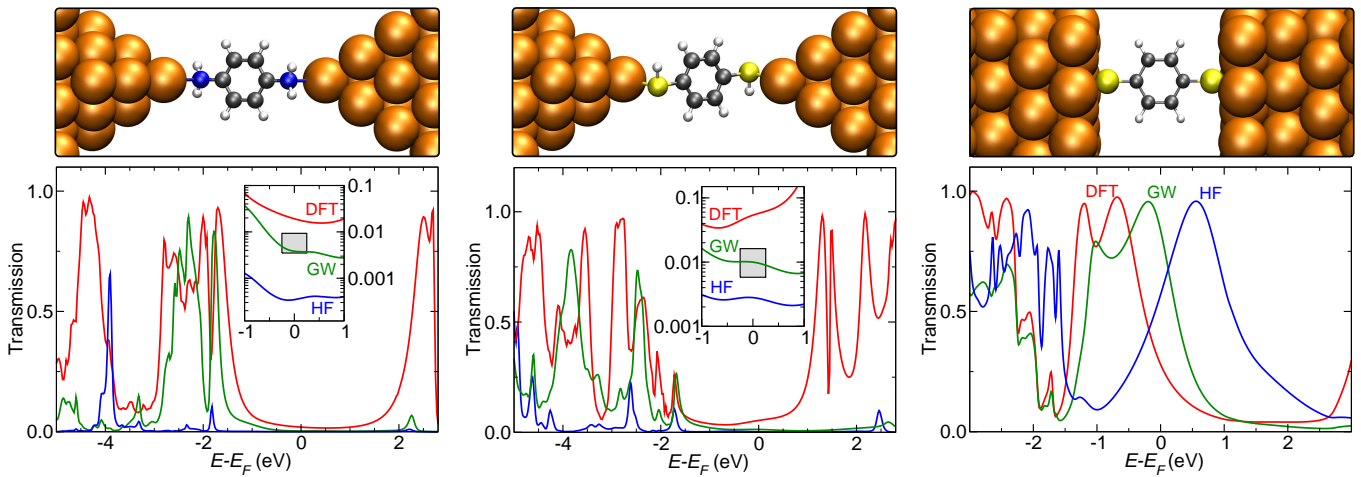}
    \caption[cap.Pt6]{\label{fig1} (Color online) Atomic structure of the BDA (left), BDT+H (middle), and BDT (right) molecular junctions. The lower panels show the transmission functions calculated using DFT-PBE (red), Hartree-Fock (blue), and the selfconsistent GW approximation (green). The insets show a zoom of the
transmission functions around the Fermi energy (set to zero). The grey boxes indicate the conductance windows
$0.007G_0 \pm 50\%$ and $0.01G_0 \pm 50\%$ which cover the the experimental values
reported in Refs. \onlinecite{venkataraman,kiguchi} and \onlinecite{song,xiao,tsutsui,kiguchi}, respectively.}
\end{figure*}

The junction geometries were optimized using the 
real space projector augmented wave method GPAW\cite{ref29}. We used a
grid spacing of 0.2~\AA~and the PBE functional for exchange and correlation\cite{ref25}. The molecules were attached to Au(111)
surfaces, modelled by a seven layer thick $4\times 4$ slab, either
directly (in the case of BDT) or via tips (in the case of BDT+H and BDA) as shown in Fig. \ref{fig1}. The surface
Brillouin zone was sampled on a $4\times 4$ $k$-point grid, and the
structures including molecule, Au tips, and outermost Au surface layers were relaxed until the residual force was below 0.05 eV/\AA. The three structures are shown in the upper panel
of Fig. \ref{fig1} and some key bond lengths are given here\cite{note}.

It is generally accepted that the hydrogen atoms dissociate from the
thiol end groups forming a gold-thiolate structure\cite{bdt_s}.
Nevertheless, our total energy calculations show that the
benzenedithiol structure has a slightly lower energy than the
benzenedithiolate when inserted between two gold tips as shown in Fig.
\ref{fig1}(middle). In these calculations the hydrogens were either
taken to be in the gas-phase or are adsorbed on the Au tips. In both
cases the energy gain is less than 0.1 eV at the equilibrium distance
but increases to 0.3 eV for a junction stretched by 1~\AA. We stress
that our calculations do not include effects of entropy which becomes
relevant at finite temperature, and furthermore assumes that hydrogen
in the gas phase is the proper reference for the solvated proton and
an electron in the electrode, i.e.  the reaction
$\text{H}^+$+$\text{e}^- \leftrightarrow \text{H}_2(\text{g})$ is in
equilibrium (in electrochemistry language we assume the standard
hydrogen potential at pH=0). For these reasons our calculations are
not sufficient to address the relative stability of thiols vs.
thiolates under the relevant experimental conditions.

Importantly we note that new experimental evidence for the chemical structure of the gold-thiolate interface at the Au(111)
surface\cite{ref18,ulstrup,voznyy} or at Au nanocluster
surfaces\cite{ref19, ref20} has recently emerged, pointing to the existence of
polymeric SR-Au(I)-SR units where the formally oxidised Au(I) adatoms
are chemically bound to thiolates and form a part of a more complex
structure (see Ref. \onlinecite{ref21} and references therein).  These complexes
are currently too challenging to treat satisfactorily at the GW
level. However, we have found that the
electronic structure of such complexes is quite similar to that of the
dithiol structure, see Sec. \ref{sec.thiol}. This is because the
hydrogen atoms play a role similar to that of the Au adatom in
passivating the sulfur atoms. Therefore the transport properties of the
dithiol structure should also be similar to those of the SR-Au(I)-SR units.

\subsection{Energy levels of isolated molecules}
A natural requirement for a method intended to describe the
energy levels of molecules in contact with electrodes, is that it
should be able to describe the energy levels of isolated molecules. As
we show below, the DFT approach fails completely in this respect
underestimating the HOMO-LUMO gaps of the three molecules by 5-6 eV
while GW energies lie within 0.5 eV of the target values.

The gas-phase molecular structures have
been relaxed in a 16~\AA~cubic cell using GPAW grid calculations as described in the previous
section. For consistency, all energy levels have been calculated using
the same double-zeta (DZ) atomic orbitals basis set. This is the same basis as used for the molecules in the transport calculations presented in the next section. For the DFT and
Hartree-Fock calculations we have found that the energies obtained
with the DZ basis agree with accurate grid calculations to within 0.2
eV. For the GW calculations the energies are within 0.1 eV of those obtained with a DZ+polarization basis set.

Table \ref{table.gas_levels} summarizes the results for the HOMO and
LUMO orbital energies obtained from the DFT-PBE eigenvalues,
selfconsistent Hartree-Fock, and selfconsistent GW. The energy levels
have been identified as the peaks in the spectral function
$\text{Tr}[\text{Im}G^r(E)]$ extrapolated to $\eta=0^+$.  Due to lack
of accurate experimental data we have also performed DFT-PBE total
energy calculations for the neutral, cation, and anion species to
obtain the addition/removal energies (last column). This approach has
been shown to produce very accurate estimates of the experimental
ionization and affinity levels of small molecules\cite{rostgaard}.

Relative to this reference, the DFT eigenvalues underestimate the
HOMO-LUMO gap of all three molecules by $5-6$~eV, Hartree-Fock
overestimates it by $2-3$~eV, while the gap obtained with GW lies
within $0.3$~eV.  These trends are consistent with a recent study of
ionization potentials of a large number of molecules\cite{rostgaard}.
The main reason for the large underestimation of the gap by DFT is the
presence of self-interactions in the PBE functional.\cite{sanvito} On
the other hand Hartree-Fock is self-interaction free; here, by virtue
of Koopmans' theorem, the overstimation of the gap can be seen as a
result of neglect of orbital relaxations. The effect of the latter is
included in GW via the screened interaction and this reduces the gap
relative the unscreened Hartree-Fock result. 

We furthermore note that DFT
places the LUMO of BDA and BDT+H below the vacuum level thus incorrectly
predicting the anions to be stable. For BDT, the LUMO level is
predicted to be negative by all methods indicating the radical nature
of this species. We note that our GW energies for BDT are in good
agreement with previous MP2 calculations\cite{mp2_BDT}.

\begin{table}
\begin{tabular} {l | l  c c c c}
\hline\hline
Molecule \hspace{0.2cm} & Orbital & DFT-PBE & \hspace{0.2cm} HF  \hspace{0.2cm}    & \hspace{0.2cm} GW \hspace{0.2cm}   & $\Delta E_{\text{tot}}$\\
\hline
BDA          & HOMO    &  -4.1   &-7.2    & -6.2  & -6.8 \\
(C$_6$H$_8$N$_2$)           & LUMO    &  -0.9   &~3.9    & ~2.9  & ~2.3 \\
         & H-L Gap  &  ~3.2   &11.1    & ~9.1  & ~9.1 \\ \hline
BDT+H         & HOMO    &  -5.1   &-8.0    & -6.9  & -7.5 \\
(C$_6$H$_6$S$_2$)           & LUMO    &  -1.3   &~3.3    & ~2.2  & ~1.3 \\
         & H-L Gap  &  ~3.8   &11.3    & ~9.1  & ~8.8 \\ \hline
BDT         & HOMO    &  -5.7   &-8.6    & -7.9  & -8.3 \\
(C$_6$H$_4$S$_2$)           & LUMO    &  -5.1   &~-1.6    & ~-2.3  & ~-2.7 \\
         & H-L Gap  &  ~0.6   &7.0    & ~5.6  & ~5.6 \\
\hline\hline
\end{tabular}
\caption{\label{table.gas_levels} Calculated frontier orbital energies of the molecules in the gas-phase. All energies are in eV and measured relative to the vacuum level. DFT-PBE refers to the Kohn-Sham eigenvalues while $\Delta E_{\text{tot}}$
  represents addition/removal energies obtained from self-consistent total energy
  calculations of the neutral, anion and cations at the DFT-PBE level. }
\end{table}

\subsection{Conductance calculations}
The transmission functions of the relaxed junction geometries were calculated as described in Sec. \ref{sec.method} using three different
approximations for $\Sigma_{xc}$, namely the PBE
xc-potential, the bare exchange potential (leading to Hartree-Fock theory), and GW. The former choice corresponds to the standard DFT-approach. All calculations employ a double-zeta basis set for the
molecules and double-zeta with polarization for the Au atoms. The results are shown in the lower
panels of Fig. \ref{fig1}, and the corresponding conductances are
summarized in Table \ref{table.conductance}. 

The conductance of BDA and BDT+H calculated with the fully
selfconsistent GW approximation agree well with the experimental
values reported in Refs. \onlinecite{venkataraman} and \onlinecite{kiguchi} for benzenediamine and
Refs. \onlinecite{song,xiao,tsutsui,kiguchi} for benzenedithiol as indicated
by the grey boxes in Fig. \ref{fig1}(left+middle).  In contrast, DFT
and Hartree-Fock respectively overestimates and underestimates the
experimental conductances by factors 5-10. Our DFT result for the BDA junction is in
good agreement with previous calculations\cite{quek,mowbray}.

It is striking that the conductance of the "classical" BDT junction
shown in Fig. \ref{fig1}(right), is predicted by all three methods to
be significantly higher than the experimental value (we obtain the
same high conductances for BDT between tips, i.e. the structure in
Fig. \ref{fig1}(middle) without hydrogen on sulfur). Our DFT
conductance is in overall good agreement with the large number of
previous calculations for the same or similar similar
structure\cite{bdt_calc}. The high conductance is clearly due to a
strong peak in the transmission function close to the Fermi level. The
peak moves to higher energies when going from DFT-PBE over GW to
Hartree-Fock, opposite to the trend normally seen for
occupied states. The peak comes from an 
unsaturated $p$-orbital on the sulphur atoms and is discussed in more detail in Sec. \ref{sec.thiol}.

These results suggest that the
structures probed in experiments on benzenedithiol junctions involve a
chemically passivated form of the thiolate linker group. As we show in
Sec. \ref{sec.thiol}, the high conductance of BDT is due to unsaturated
$p$-states on the sulphurs with energy close to the Fermi level (the
energy of this orbital does not change considerable with DFT,
Hartree-Fock and GW). In the thiol and SR-Au(I)-SR structures, these
states form bonds to H and the Au adatom, respectively, and are
thereby shifted away from the Fermi level. On the other hand, the
electronic structure and transmission functions of BDT+H and
SR-Au(I)-SR structure are rather similar indicating that the BDT+H can be
viewed as a simple model of the more complex SR-Au(I)-SR structure.

It should of course be kept in mind that experiments are performed in
solution and at room temperature and are subject to variations in the
detailed atomic structure. Thus the measured conductance values should
not be considered as highly accurate references for theoretical
calculations on idealized junctions.

\begin{table}
\begin{tabular} {l | c c c }
\hline\hline
Method               &  \hspace{0.5cm}   BDA \hspace{0.5cm}          & BDT+H                &    \hspace{0.5cm}    BDT  \hspace{0.5cm}         \\
\hline
DFT-PBE              & $2.1\cdot 10^{-2}$ & $5.4\cdot 10^{-2}$  & $2.8\cdot 10^{-1}$  \\
HF                   & $4.0\cdot 10^{-4}$ & $2.7\cdot 10^{-3}$  & $5.7\cdot 10^{-1}$  \\
GW                   & $4.2\cdot 10^{-3}$ & $1.0\cdot 10^{-2}$  & $8.3\cdot 10^{-1}$  \\
G$_0$W$_0$(PBE)      & $8.0\cdot 10^{-3}$ & $1.6\cdot 10^{-2}$  & $7.5\cdot 10^{-1}$  \\
G$_0$W$_0$(HF)       & $2.2\cdot 10^{-3}$ & $9.8\cdot 10^{-3}$  & $8.7\cdot 10^{-1}$  \\
\hline \hline
\end{tabular}
\caption{\label{table.conductance} Calculated conductance in units of $G_0$ for the three junctions shown in Fig. \ref{fig1}. The last two rows refer to non-selfconsistent GW calculations based on the DFT-PBE or Hartree-Fock Green function, respectively.}
\end{table}

\begin{figure}[!h]
    \includegraphics[width=0.9\linewidth,angle=0]{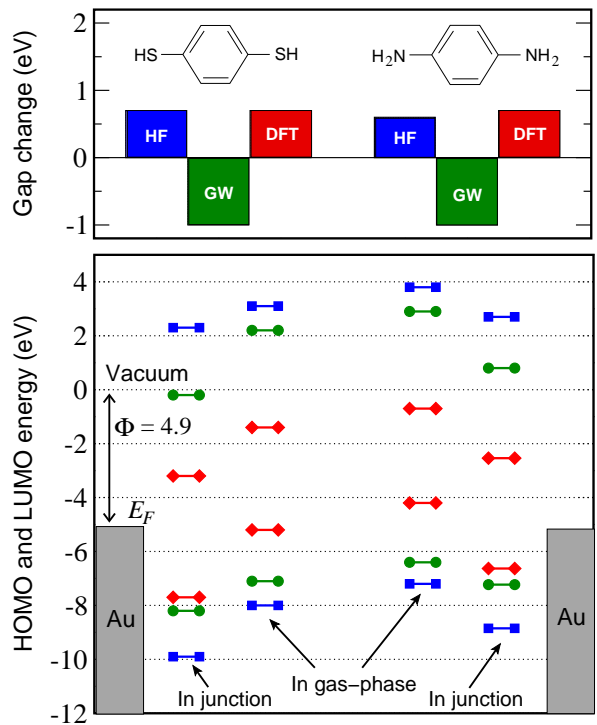}
    \caption[cap.Pt6]{\label{fig2} (Color online) Upper panel shows the change in the HOMO-LUMO gap as the molecules are brought from the gas-phase into the junction. Lower panel shows the energy of the HOMO and LUMO levels (relative to vacuum) in the gas phase and in the junction. The left and right set of levels correspond to BDT+H and BDA, respectively. The calculated work
function of the gold junction (with tips) in absence of a molecule is $\Phi=4.9$ eV as indicated.}
\end{figure}

\subsection{Energy level alignment}
In Fig. \ref{fig2} we show the calculated HOMO and LUMO energy levels
of BDT+H and BDA in the gas-phase and in the junction. All energies
have been aligned relative to the vacuum level. At this point we note
that an accurate description of the vacuum level, i.e. the work
function, can in general be difficult to obtain with an atomic orbital
basis\cite{lorente}.  However, by using more diffuse basis functions (an energy shift of 0.01 eV has been used for all Au basis functions throughout this work\cite{GPAW-LCAO}) we obtain a work function for Au(111) of 5.4 eV
in good agreement with the experimental value of 5.31
eV\cite{au_work}. At the position of the molecule, i.e. in the region
between the two tips, the electrostatic potential from a calculation
where the molecule has been removed, converges to a constant value of
4.9 eV above the metal Fermi level. This value has been used as
reference for the vacuum level in Fig. \ref{fig2}. In the junction
where the levels are broadened due to hybridization with the metal
states, the level positions have been defined as the first moment of
the projected density of states of the relevant molecular orbital,
$\text{Im}\langle \psi_n|G^r(E)|\psi_n \rangle$. Here the
$|\psi_n\rangle$ are obtained by diagonalizing the KS Hamiltonian
within the molecular subspace.

The orbital energies obtained from a GW calculation include the
dynamical response of the electron system to the added electron/hole
via the correlation part of the self-energy. In general the
correlations will shift the occupied levels up and the empty levels
down relative to the bare Hartree-Fock energies. When a molecule is
brought from the gas-phase into a junction the electronic character of
its environment changes from insulating to metallic. The enhanced screening should thus
cause the gap to shrink (neglecting shifts due to hybridization) as compared to its gas-phase value.
However, it has recently been shown for molecules weakly bonded to a
metal surface, that this effect is completely missing in effective
single-particle theories based on a (semi)local description of
correlations \cite{neaton,juanma}. 

As a result, in our DFT and
Hartree-Fock calculations, the change in the frontier orbital energies
induced by the coupling to gold is completely governed by the effect
of hybridization which tends to open the gap by $~0.7$~eV for both
molecules, see top panel of Fig. \ref{fig2}. In contrast, the GW gap
is reduced by $1$~eV due to the enhanced screening in the contact.
Since the hybridization shift is of course also present in he GW
calculation, we conclude that the enhanced screening due to the metal
contacts reduces the HOMO-LUMO gap by $~1.7$~eV relative to the value
in the gas-phase. 

Note that we refrain from using the term "image charge effect" to
describe the renormalization of molecular orbitals. This term is
appropriate for weakly coupled molecules where the screening of the
added electron/hole takes place within the metal. For intermediate or
strongly coupled molecules, there is no clear distinction between
metal states and molecular states, and the screening is more
appropriately described as dynamical charge
transfer\cite{thygesen_image}.

From Fig. \ref{fig2} it follows that the HOMO level of the molecules
in the junction is predicted by DFT-PBE to lie $0.5-0.7$~eV higher
than obtained with GW. This agrees well with a recent study of
benzenediamine derivatives on gold(111) which showed that DFT places
the HOMO level about 1 eV too high compared to UPS
measurements\cite{ref3}. The fact that the DFT-PBE description of the
energy levels is better for the adsorbed rather than isolated
molecules may be seen as a result of the metallic screening build into
the DFT xc-functional via its origin in the homogeneous electron
gas\cite{rohlfing}. It should also be noted that the error (compared to GW) of the
DFT eigenvalues is significantly larger for the LUMO than for the
HOMO. This is in good agreement with previous plane wave calculations
for molecule/metal interfaces\cite{juanma}.

\subsection{G$_0$W$_0$ calculations}
\begin{figure}[!h]
    \includegraphics[width=1.0\linewidth,angle=0]{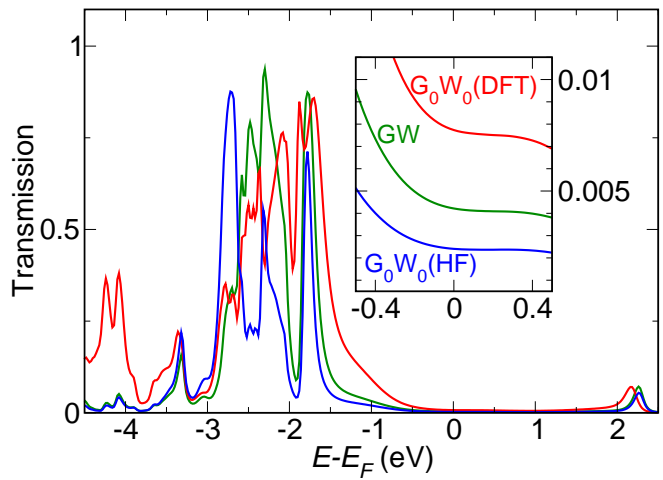}
    \caption[cap.Pt6]{\label{figg0w0} (Color online) Transmission functions for the BDA junction calculated with selfconsistent GW and non-selfconsistent G$_0$W$_0$ using either the Hartree-Fock or the DFT-PBE Green functions as input.}
\end{figure}

To test the role of selfconsistency (in the GW and Hartree self-energies) we have performed
non-selfconsistent G$_0$W$_0$ calculations using both the DFT-PBE
Green function and the (selfconsistent) Hartree-Fock Green function as
the initial $G_0$. The results for the conductance of all systems are
summarized in the last two rows of Table \ref{table.conductance}. We
notice is that the conductance can vary by a factor of three depending
on $G_0$. We also note that the Hartree-Fock starting point comes closer to the selfconsistent result. This is because the Hartree-Fock Green function is an overall better approximation to the GW Green function than is the DFT Green function (see e.g. the comparison of frontier orbital energies in Fig. \ref{fig2}). As an example Fig. \ref{figg0w0} shows the calculated transmission
functions for the BDA junction in a larger energy range around
$E_F$.

For the BDA and BDT+H junctions, G$_0$W$_0$[PBE] overestimates the conductance while G$_0$W$_0$[HF] underestimates the conductance relative to GW.
This can be understood as
follows: Since DFT-PBE (Hartree-Fock) underestimates (overestimates)
the HOMO-LUMO gap, the use of these Green functions to evaluate the GW
self-energy will lead to an overestimation (underestimation) of the
screening. As discussed above the screening contained in the correlation part of the GW self-energy tend to
reduce the HOMO-LUMO gap. This reduction of the gap is thus
overestimated in the G$_0$W$_0$[PBE] calculation and underestimated in the
G$_0$W$_0$[HF] calculation which explains the trends in conductance, i.e.
more screening $\to$ smaller gap $\to$ higher conductance. 

The G$_0$W$_0$ results
for the strongly coupled BDT junction show less variation. This is
perhaps surprising given the large difference between the DFT-PBE and
Hartree-Fock results shown in Fig. \ref{fig1}. However, DFT-PBE and
Hartree-Fock give almost equal density if states at the Fermi level
(the transmission functions at the Fermi energy are also rather
similar), and therefore the screening contribution obtained with the
two choices for G$_0$ is also very similar.

\subsection{Scissors operator}
In this section we investigate to what extent the GW transmission function can be
reproduced by a DFT calculation where the occupied and unoccupied
molecular levels have been shifted rigidly to match the GW energies. This is illustrated by applying a scissors
operator (SO) to the Au-BDA-Au junction shown in Fig. \ref{fig1}(left). After shifting the molecular levels we
perform a non-selfconsistent calculation of the transmission function.
A similar procedure has previously been successfully used to include
image charge effects and correct for self-interaction errors in
DFT-transport calculations\cite{quek,mowbray,quek2}. We refer the reader to Ref. \onlinecite{mowbray} for
more details on the SO technique applied here.

\begin{figure}[!h]
    \includegraphics[width=1.0\linewidth,angle=0]{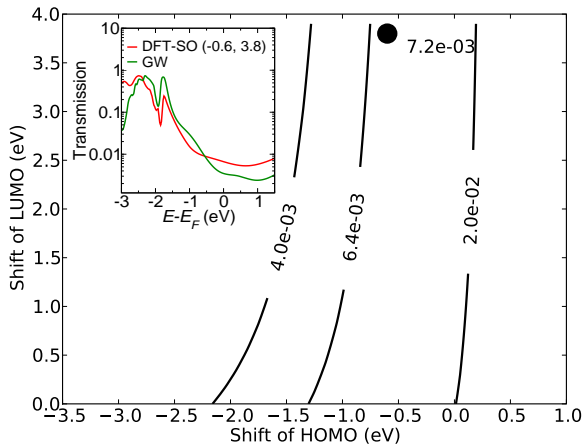}
    \caption[cap.Pt6]{\label{fig5} (Color online) DFT conductance of the Au-BDA-Au junction for different shifts of the occupied and unoccupied molecular orbital energies. The three iso-contours with values of 4.2x10-3, 6.4x10-3 and 2.1x10-2  in units of $G_0$ correspond to the GW, experimental and DFT conductance values, respectively. The dot indicates the conductance obtained by fitting the DFT HOMO and LUMO level positions to the corresponding GW level positions. Inset: Transmission function calculated with GW (green) and DFT+SO (red). The SO shifts are -0.6 eV and 3.8 eV for the occupied and unoccupied molecular orbital, respectively, which makes the HOMO and LUMO energies of the DFT calculation coincide with the GW energies.}
\end{figure}
 
In Fig. \ref{fig5} we show contour plots of the DFT conductance for the BDA
junction where the shift of the occupied and unoccupied molecular
orbital energies has been varied independently over 4 eV.  The three
values for the contour lines shown correspond to the conductance
obtained with GW (0.0042G$_0$), the experimental conductance (0.0064G$_0$)
and the DFT conductance (0.021G$_0$), respectively. The dot indicates
the energy shifts which make the HOMO and LUMO levels of the DFT
calculation match the corresponding levels of the GW calculation; the
required SO shifts are -0.6eV and 3.8eV for the occupied and
unoccupied molecular orbitals, respectively. Shifting the levels by
this amount reduces the DFT conductance by a factor of ~3 from 0.02G$_0$
to 0.0072G$_0$. Interestingly, the GW conductance is not reproduced by
these shifts; it is even lower by a factor of ~1.5. In fact, to
reproduce the GW conductance a shift of about -1.3eV of the DFT HOMO
is required (keeping the LUMO position fixed at the GW position). This
shows that while the renormalization of the molecular level energies
can explain the main part of the difference between the DFT and GW
conductance, the different shape of the transmission resonances also
plays a role. This is clear from the inset which shows the GW transmission function (green) and the DFT transmission with SO chosen to match the GW HOMO and LUMO levels (the dot in the main figure). The lower conductance obtained with GW is seen to be a consequence of a faster decay of the HOMO resonance towards the Fermi
level. This difference comes from the energy dependence of the GW self-energy.

\subsection{Thiol vs. thiolate structures}\label{sec.thiol}
\begin{figure}[!h]
    \includegraphics[width=0.9\linewidth,angle=0]{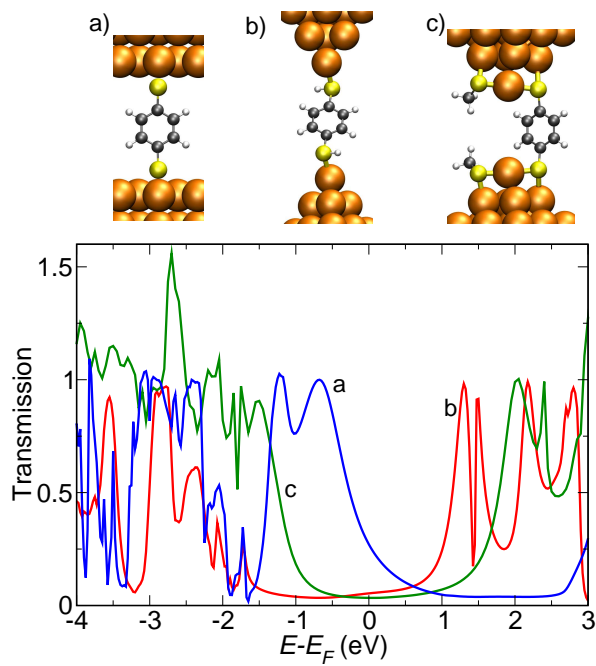}
    \caption[cap.Pt6]{\label{figabc} (Color online) DFT-PBE transmission functions for three different structures of the Au-BDT-Au junction: a) the "classical" benzenedithiolate, b) benzenedithiol, and c) the SR-Au(I)-SR complex (with two benzene molecules replaced by $\text{CH}_3$ units for simplicity). The transmission function of b) and c) are rather similar while that of a) shows a peak close to the Fermi level.}
\end{figure}

In Fig. \ref{figabc} we compare the DFT transmission
functions for: (a) the ``classical'' structure of benzenedithiolate
[structure in Fig. \ref{fig1}(right)] (b) benzenedithiol between tips
on Au(111) [structure in Fig. \ref{fig1}(middle)], and (c)
benzenedithiolate in a SR-Au(I)-SR molecular unit form [structure 2 of
Fig. 1 in Ref. \onlinecite{ref21}].  The conductance (essentially the
transmission function at the Fermi energy) obtained for structures (b)
and (c) is very similar but markedly different from (a). The high
conductance of the structure (a) is due to the strong transmission
resonance lying just below the Fermi level. This resonance has also
been found in many previous studies and seems to be a characteristic
and robust feature of this junction. As shown in this work this
transmission resonance is also present in GW and Hartree-Fock
calculations. In contrast, for both structure (b) and (c) the transmission function is rather flat in an energy window of $\pm 1$ eV around the Fermi level.

\begin{figure}[!h]
    \includegraphics[width=0.9\linewidth,angle=0]{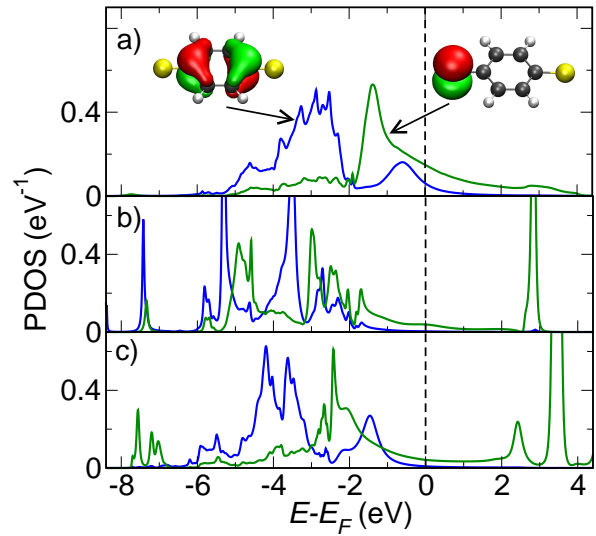}
    \caption[cap.Pt6]{\label{figgroup} (Color online) Projected
      density of states for the HOMO-1 of the C$_6$H$_4$ moeity and
      its group orbital (sulfur $p$ orbital). These orbitals essentially determine the transmission function around the Fermi energy for all the three structuers.
It should be noticed
      that for structure (a) (the "classical" BDT structure) the sulfur
      $p$-orbital has a peak in the PDOS just below the Fermi level
      which is responsible for the high conductance. In (b) and (c) the
      sulfur is chemically passivated and the PDOS of the $p$-orbital
      splits into bonding and anti-bonding states at $\pm 3$ eV thereby lowering the
      conductance.}
\end{figure}

To examine the origin of the different transmission functions found
for structures (a) and (b,c) we found it useful to consider the
molecular orbitals of the C$_6$H$_4$ moiety of the three molecules. By
diagonalizing the (Kohn-Sham) Hamiltonian of this part of the
molecules we found that the orbital depicted to the left in the upper
panel of Fig. \ref{figgroup}, and which constitute the HOMO-1 of the
C$_6$H$_4$ moiety, is responsible for all the structure in the
transmission functions below the Fermi
level.  Note, that the orbitals obtained in this way are different
from the HOMO and LUMO levels shown in Fig. \ref{fig2} which were
obtained by diagonalizing the Hamiltonian for the entire molecules
including the SH and NH$_2$ end groups.

The different panels of Fig. \ref{figgroup} show the projected density
of states (PDOS) of the HOMO-1 for the three structures together with
the PDOS of the sulfur $p$-orbital to which the HOMO-1 couples. Within
the Newns-Anderson model\cite{newns}, the sulphur $p$-orbital is the
so-called group orbital.  Note, that the PDOS of the $p$-orbital has been calculated in the absence of coupling to the HOMO-1 as it should in the Newns-Anderson model. The transmission function is then essentially
the product of the PDOS of the HOMO-1 and the PDOS of the group
orbital\cite{wannier_thygesen}.  It is clear that the origin of the
(double) transmission peak around -1 eV for structure (a) is due to a
resonance formed by the HOMO-1 and the sulfur $p$-orbital. The
chemical passivation of sulfur, by either hydrogen or the Au adatom,
implies that the PDOS of the $p$-orbital splits into bonding and
anti-bonding states around -3 eV and 3 eV, respectively.  This in turn
shifts the PDOS of the HOMO-1 down in energy. In particular its
magnitude around the Fermi level is lowered and as a consequence the
transmission function (being essentially the product of the two
curves) is suppressed in a broad energy window around $E_F$. Thus
chemical passivation of the sulfur is the main reason for the lower
conductance observed in structures (b,c) as compared to (a). A
secondary effect, giving rise to differences in the transmission of
structure (b) and (c) is the different interface dipoles which shift the
electrostatic potential at the C$_6$H$_4$ moiety by different amounts. This shift, however, has little influence on the conductance due to the flatness of the transmission functions around $E_F$. 

To verify this scenario, we have applied a scissors operator of 1 eV
to the C$_6$H$_4$ moiety of structure (b) in order to align the
onsite energy of the HOMO-1 to that of structure (c). The
resulting transmission function (not shown) is very similar to that of (c) and the conductance is $3.5\cdot 10^{-2}G_0$ compared to $3.3\cdot 10^{-2}G_0$ obtained for structure (c).
On the other hand, the conductance of BDT cannot be brought below $0.1G_0$ by
sifting the levels of the C$_6$H$_4$ moiety down (by up to 2 eV) because of the strong peak
around -1 eV due to the sulphur $p$-orbital.

Although the above picture is based on the DFT electronic structure,
the qualitative similarities of the DFT and GW transmission functions
in Fig. \ref{fig1} implies that the same picture applies to the GW calculations.

\section{Summary}
We have presented a first-principles method for modelling quantum
transport in molecular nanostructures beyond the single-particle
approximation. The method is based on non-equilibrium Green functions
and applies to the general case of a finite bias voltage, but in this work we
focused on the zero bias regime.  The conductance of
benzenedithiolate (BDT), benzenedithiol (BDT+H), and benzenediamine
(BDA) was calculated using the selfconsistent GW approximation. In
contrast to standard DFT and Hartree-Fock methods, the GW
approximation was found to yield consistently accurate values for the
energy levels of both isolated and contacted molecules due to its proper treatment of self-interaction and
dynamical screening. In general, results obtained with GW for the
electronic conductance and energy gaps of contacted molecules lie in
between the values obtained with DFT and Hartree-Fock. The latter
methods respectively overestimates and underestimates the screening
and none of them are able to detect the change in the molecule's
electronic environment when it is coupled to electrodes.

Non-selfconsistent G$_0$W$_0$ calculations were found to yield
conductance values within 50\% of the GW results depending on the
initial G$_0$. It was shown that the main difference between the GW
and DFT calculations comes from the renormalization of the position of the
molecular energy levels. However, changes in the shape of
transmission resonances, and thus the conductance, also occur due to the energy-dependence of the
GW self-energy.

The GW conductance of benzenediamine was found to be in good agreement with
experiments. For benzenedithiolate between Au(111) surfaces we found a
conductance well above $0.1G_0$ and thus conclude that this structure
cannot be responsible for the measured conductance around 0.01$G_0$.
On the other hand, a conductance close to this value was found for a
hydrogenated benzenedithiol junction which, as we demonstrated, represents
a reasonable model of more complex gold/thiolate structures where the
chemical passivation of sulfur is provided by a gold adatom rather
than hydrogen. These gold/thiolate structures are presently too
demanding for our selfconsistent GW method, however, our results
suggest that such structures are responsible for the
0.01$G_0$ conductance reported experimentally.

In conclusion, we showed that a consistent and quantitatively accurate description of
energy level alignment and charge transport in phase-coherent molecular conductors can be
obtained from first-principles with the (computationally feasible) GW approximation.
We believe this development is important for increased the synergy between theory and experiments in molecular electronics which is essential for continued progress in the field.

\section{Acknowledgements}
We would like to thank Angel Rubio and Karsten Jacobsen for inspiring discussions.
The authors acknowledge support from the Lundbeck Foundation's Center for
Atomic-scale Materials Design (CAMD), the Danish Center for Scientific Computing and
the Academy of Finland. Part of the computations were done at the CSC - the Finnish
IT Center for Science in Espoo.

\appendix
\section{The GW self-energy}
\label{sec:GWselfenergy}

Let $U$ denote the rotation matrix that diagonalizes the pair orbital
overlap $S_{ij,kl} = \braket{n_{ij}}{n_{kl}}$, i.e. $U^\dagger S U =
\sigma I$. The columns of $U$ are truncated to those which have
corresponding eigenvalues $\sigma_q > 10^{-5} a_0^{-3}$. We then only
calculate the reduced number of Coulomb elements
\begin{equation}
  v_{qq'} = \bra{n_q} \frac{1}{|\bold r-\bold r'|} \ket{n_{q'}},
\end{equation}
where $n_q(\bold r)$ are the optimized pair orbitals
\begin{equation}
  n_q(\bold r) = \sum_{ij} n_{ij}(\bold r) U_{ij,q} / \sqrt{\sigma_q},
\end{equation}
which are mutually orthonormal, i.e. $\braket{n_q}{n_{q'}} =
\delta_{qq'}$.

The determination of the GW self-energy proceeds by calculating first the lesser and greater components of the 
polarization matrix in the time domain
\begin{align}
  P^<_{ij,kl}(t) &= 2i G^<_{ik}(t) G^{>}_{jl}(t)^*,\label{eq:P-lesser}\\ 
 P^>_{ij,kl}(t) &= P^{<}_{ji,lk}(t)^*.
\end{align}
where the factor 2 accounts for spin and we have used $G^{>}_{lj}(-t)=-G^{>}_{jl}(t)^*$.
This is then downfolded to the reduced representation
\begin{equation}
  P^{\lessgtr}_{qq'} = \sum_{ij,kl} \sqrt{\sigma_q} U^*_{ij,q} P^{\lessgtr}_{ij,kl}
U_{kl,q'} \sqrt{\sigma_{q'}}.
\end{equation}
The screened interaction can be determined from the lesser and greater
polarization matrices, and the static interaction $v_{qq'}$, via the
relations
\begin{align}
  P^r(t) &= \theta(t) \left( P^>(t) - P^<(t) \right),\\
  W^r(\omega) &= [I - v P^r(\omega)]^{-1}v,\\
  W^>(\omega) &= W^r(\omega) P^>(\omega) W^{r}(\omega)^{\dagger},\\
  W^<(\omega) &= W^>(\omega) - W^r(\omega) + W^{r}(\omega)^{\dagger},
\end{align}
where all quantities are matrices in the optimized pair orbital basis
and matrix multiplication is implied. We obtain the screened
interaction in the original orbital basis from
\begin{equation}
  W^{\lessgtr}_{ij,kl}(\omega) \approx \sum_{qq'} U_{ij,q} \sqrt{\sigma_q}
W^{\lessgtr}_{qq'}(\omega) \sqrt{\sigma_{q'}} U_{kl,q'}^*,
\end{equation}
which is an approximation due to the truncation of the columns of $U$.
Finally the GW self-energy can be determined from
\begin{align}
  \Sigma^{\lessgtr}_{\text{GW,}ij}(t) &= i \sum_{kl} G^{\lessgtr}_{kl}(t)
W^{\lessgtr}_{ik,jl}(t)\\
  \Sigma^r_\text{GW}(t) &= \theta(t) \left( \Sigma^>_\text{GW}(t) -
\Sigma^<_\text{GW}(t) \right)+\delta(t)\Sigma_x.
\end{align}
The exchange and Hartree potentials are given by
\begin{align}
  \Sigma_{x,ij} &= i\sum_{kl}v_{ik,jl}G^<_{kl}(t=0)\\
  V_{\text{H},ij} &= -2i\sum_{kl}v_{ij,kl}G^<_{kl}(t=0) 
\end{align}
The latter equals the first term in Eq. (\ref{eq.hartree}).
 
In equilibrium the lesser and greater Green functions are obtained from  
\begin{align}
  G^<(E)&=-f(E-\mu)[G^r(E)-G^r(E)^\dagger]\\
  G^>(E)&=(1-f(E-\mu))[G^r(E)-G^r(E)^\dagger]\label{eq:g-greater}
\end{align}
where $f$ is the Fermi-Dirac distribution function. For self-consistent calculations, equation
\eqref{eq:P-lesser}-\eqref{eq:g-greater} are iterated untill
convergence in $G$.

\section{Size of the "extended molecule"}
\label{app3}
\begin{figure}[!h]
    \includegraphics[width=0.9\linewidth,angle=0]{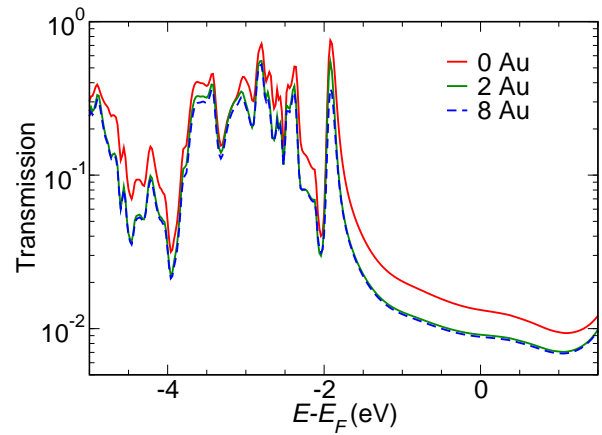}
    \caption[cap.Pt6]{\label{figconv} (Color online) Transmission function for the BDA junction calculated at the G$_0$W$_0$(PBE) level and with a minimal single-zeta basis set. The three curves correspond to different number of gold atoms in the extended molecule region.}
\end{figure}

As explained in Sec. \ref{sec.gw} the GW self-energy of the molecule
has contributions from screening diagrams extending into the 
electrodes. This is included by calculating $P$ and $W$ for the
extended molecule which includes the Au atoms closest to the molecule.
In Fig. \ref{figconv} we show the dependence of the transmission
function on the size of the extended region. The latter has been
varied from zero Au atoms (extended molecule coincides with molecule),
to the Au tip atoms, to the four-atom Au pyramids on each side of the
molecule. From this we conclude that it is sufficient to include
polarization diagrams for the Au tip atoms as was done for the
calculations presented in the main text. We mention that, for
simplicity, the calculations shown in Fig. \ref{figconv} have been performed
at the G$_0$W$_0$(PBE) level and using a minimal single-zeta basis
set. However, the conclusions regarding the convergence of the GW
self-energy with respect to the size of the extended molecule, should
be equally valid for the case of selfconsistent GW with the DZ/DZP basis set used for all calculations in the main text.

\bibliographystyle{unsrt}

\begin{thebibliography}{100}
\bibitem{note}
For BDA the distance between the Au tip atoms was 9.90~\AA, and the
  Au-N bondlength was 2.55~\AA. For BDT+H the distance between the Au tip atoms was 9.80~\AA~and the Au-S bondlength was 2.47~\AA. The BDT was adsorbed on the Au(111) surface in a fcc-hollow
  position slightly tilted towards the bridge site. The distance
  between the Au(111) surfaces was 10.20~\AA.
\end{thebibliography}

\end{document}